\documentclass{article}
\usepackage{spconf,amsmath,amssymb,graphicx,adjustbox,hyperref}
\usepackage{color,xcolor}
\usepackage[hang,flushmargin]{footmisc} 
\usepackage{siunitx}
\usepackage{booktabs}
\usepackage{caption}
\usepackage{fancyhdr}
\usepackage{multirow}
\usepackage{url}
\usepackage{cite}

\title{CosyAccent: Duration-Controllable Accent Normalization Using Source-Synthesis Training Data}

\name{
\textit{Qibing Bai}$^{1,2,5}$, 
\textit{Shuhao Shi}$^{1}$, 
\textit{Shuai Wang}$^{4,6\dag}$,
\textit{Yukai Ju}$^5$,   
\textit{Yannan Wang}$^5$,   
\textit{Haizhou Li}$^{1,2,3,6}$\thanks{$\dag$: Corresponding author} 
}
\address{
\textsuperscript{1}SDS, \textsuperscript{2}SRIBD, and \textsuperscript{3}SAI, The Chinese University of Hong Kong, Shenzhen, China\\
\textsuperscript{4}School of Intelligence Science and Technology, Nanjing University, Suzhou, China \\
\textsuperscript{5}Tencent Ethereal Audio Lab, Tencent, Shenzhen, China \\
\textsuperscript{6}Shenzhen Loop Area Institute, Shenzhen, China \\
}

\begin{document}
\ninept
\maketitle
\begin{abstract}
Accent normalization (AN) systems often struggle with unnatural outputs and undesired content distortion, stemming from both suboptimal training data and rigid duration modeling. In this paper, we propose a ``source-synthesis" methodology for training data construction. By generating source L2 speech and using authentic native speech as the training target, our approach avoids learning from TTS artifacts and, crucially, requires no real L2 data in training.
Alongside this data strategy, we introduce CosyAccent, a non-autoregressive model that resolves the trade-off between prosodic naturalness and duration control. CosyAccent implicitly models rhythm for flexibility yet offers explicit control over total output duration. Experiments show that, despite being trained without any real L2 speech, CosyAccent achieves significantly improved content preservation and superior naturalness compared to strong baselines trained on real-world data.\footnote{Samples: {\url{https://p1ping.github.io/CosyAccent-Demo}}. Code \& data: {\url{https://github.com/P1ping/CosyAccent}}.}
\end{abstract}
\begin{keywords}
Accent conversion, speech synthesis, voice conversion, duration control
\end{keywords}

\section{Introduction}
\label{sec:intro}

Accent normalization (AN)\footnote{Also referred to as Foreign Accent Conversion (FAC).} seeks to remove the accent from a non-native (L2) speech to obtain a native sounding (L1) speech while preserving the speaker's unique vocal identity and the original linguistic content. The practical applications of this technology are wide-ranging, from improving pronunciation for language learners~\cite{felps2009foreign} to enhancing the authenticity of dubbed media~\cite{turk2002subband} and enabling personalized text-to-speech (TTS) systems~\cite{sun2016personalized}.
A primary challenge in accent normalization is the scarcity of large-scale, parallel L1-L2 corpora where the same speaker utters the same content in two different accents. This data bottleneck has forced the field to evolve, moving from early methods that required reference speech during inference~\cite{zhao2018icassp, zhao2019foreign, li2020improving, ding2022accentron} or were trained on limited content-paired data~\cite{ liu2020end,jia2024convert}. To overcome these constraints, the dominant paradigm has shifted towards the use of synthetic data to construct training pairs.

Current synthetic data strategies predominantly focus on \textit{target synthesis}, which generates L1 speech using TTS models and trains a conversion model on these synthetic pairs~\cite{zhou2023tts,chen2024transfer,nguyen2024improving, bai2024diffusion}. Another strategy involves synthesizing \textit{both} the source and target speech using multi-accent TTS models, often leveraging discrete representations to mitigate error accumulation~\cite{nguyen24_syndata4genai}. Other methods have used voice conversion to expand small L2 corpora~\cite{zhao2021converting,nguyen2022accent} or transliteration to create pairs~\cite{inoue2025macst}, though these often face limitations in scale.
However, these methods share a vulnerability: the model's quality is capped by the authenticity of the synthesized L1 target. Artifacts or unnatural prosody from the TTS system are inherited by the AN model, creating a performance ceiling that persists even when using intermediate SSL token representations~\cite{nguyen24_syndata4genai,bai25_interspeech}.

In this paper, we propose to synthesize the speech in a different way by synthesizing the \textit{source} L2 speech from a large-scale, high-quality native L1 corpus. This ``source-synthesis" approach ensures that our model is trained on genuine, high-quality native speech as its target, with natural prosody and voice quality, and without artifacts that come with target-synthesis methods. 
Furthermore, in contrast to dual-synthesis methods that require fine-tuning a TTS model on limited L2 data~\cite{nguyen24_syndata4genai}, our method leverages a powerful, prompt-based TTS~\cite{du2024cosyvoice}. This makes our data generation process highly scalable and much less dependent on real L2 data collection.

Together with the new data strategy, we also propose a novel model, \textit{CosyAccent}. It is a non-autoregressive (NAR) system designed to resolve the common trade-off between prosodic flexibility and duration control. Unlike frame-to-frame methods~\cite{jin2023voice,chen2024transfer,bai2024diffusion,nguyen2024improving} that rigidly copy source timing or sequence-to-sequence models that lack control, CosyAccent implicitly models rhythm for naturalness while offering explicit control over the total output duration, making it adept for tasks like dubbing where the duration of output speech needs to be preserved.
Our contributions are three-fold:
\begin{itemize}
    \item We introduce a novel data generation strategy that synthesizes L2 source speech from high-quality L1 corpora, eliminating the dependency on real accent data while ensuring training on authentic, artifact-free native targets.

    \item We propose CosyAccent, a NAR accent normalization model capable of generating high-quality speech with a specified total duration, resolving the trade-off between prosodic flexibility and duration control for applications like dubbing.

    \item We empirically show that, by training exclusively on our source-synthesized data, our model achieves superior content preservation and naturalness, matching or exceeding strong baselines that were trained on real-world accented data.
\end{itemize}

\section{Related-Work}
\label{sec:related}

\subsection{Synthetic Data for Accent Conversion}
The scarcity of parallel L1-L2 data has driven the field towards synthetic data generation. The dominant strategy is \textit{target-synthesis}, where a TTS model generates the L1 target speech~\cite{chen2024transfer,nguyen2024improving,bai2024diffusion,bai25_interspeech}. However, this approach is limited, as any TTS artifacts are inherited by the AN model. To mitigate this, some methods synthesize \textit{both} source and target speech~\cite{nguyen24_syndata4genai}, often using discrete tokens to reduce error accumulation~\cite{nguyen24_syndata4genai,bai25_interspeech}. Other approaches, such as using voice conversion~\cite{nguyen2022accent} or transliteration~\cite{inoue2025macst}, often struggle to scale.

In contrast, our source-synthesis approach inverts this paradigm. We use a powerful prompt-based TTS, CosyVoice2~\cite{du2024cosyvoice}, to generate the L2 source, ensuring the model trains on authentic, artifact-free L1 targets. This strategy is highly scalable and avoids the performance ceiling of target-synthesis methods.

\subsection{Duration Modeling in Speech Conversion}
Duration modeling in speech conversion presents a trade-off between the rigid timing of frame-to-frame models~\cite{chen2024transfer,nguyen2024improving} and the flexibility of sequence-to-sequence (seq2seq) models, which lack explicit duration control~\cite{huang2021any}. To bridge this gap, recent non-autoregressive methods have introduced explicit, phoneme- or token-level duration predictors~\cite{lee2022duration,kreuk2022textless,oh2025durflex}.

This is crucial for AN, as mismatched temporal patterns are a key component of an accent. While some AN models maintain the source's total duration~\cite{bai25_interspeech}, our work takes a different approach. CosyAccent implicitly models rhythm for naturalness—akin to seq2seq—while still allowing explicit control over the total output duration, offering a novel solution to this trade-off.

\section{Method}
\label{sec:method}
\subsection{Construction of Training data}
\label{subsec:data_construction}

\begin{figure}[htbp]
\centering
\vspace{-0.2cm}
\includegraphics[width=0.42\textwidth]{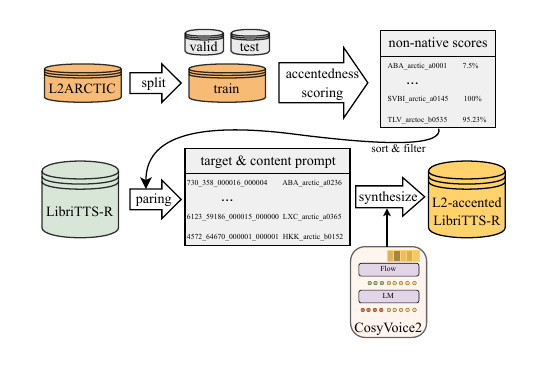}
\vspace{-0.2cm}
\caption{Construction pipeline of the paired training data.}
\vspace{-0.5cm}
\label{fig:data_pipeline}
\end{figure}

We construct our paired dataset by synthesizing L2 source audio for the native LibriTTS-R corpus~\cite{koizumi2023librittsr}. Our \textit{source-synthesis} approach method leverages CosyVoice2~\cite{du2024cosyvoice}, a dual-prompt TTS model that empirically allows separate control over speaking style and timbre. We synthesize each L2 source sample by providing two prompts: an L2 utterance from L2-ARCTIC~\cite{zhao2018l2arctic} to set the accent, and the original L1 target utterance to preserve the speaker's timbre. This process yields a synthetic L2 source that is perfectly aligned with its authentic L1 target in both content and speaker identity.
The data pipeline is illustrated in Figure~\ref{fig:data_pipeline}. It consists of four steps below.

\noindent\textbf{Subset split.} We first partition the L2-ARCTIC dataset into training, validation, and testing subsets. Since L2-ARCTIC uses the same prompt sentences across speakers, we construct the split: 50 sentences for validation and 80 for testing. So the testing sentences are not indirectly seen during training, guaranteeing a fair evaluation.

\noindent\textbf{Accentedness scoring.} To ensure high-quality L2 characteristics, we filter the L2-ARCTIC samples. We use a pre-trained accent classifier\footnote{https://huggingface.co/Jzuluaga/accent-id-commonaccent\_xlsr-en-english} to score each sample's non-native accent strength. We retain all samples with a score above 0.5. We select at least the top 200 utterances per speaker to ensure speaker diversity.

\noindent\textbf{Pairing L1 and L2 Data.} Each native utterance in LibriTTS-R is randomly assigned an L2 prompt from the filtered set. The assignment process is balanced to ensure that all L2 speakers are used a similar number of times.

\noindent\textbf{Synthesis.} In the final step, we generate the synthetic L2 speech. For each L1 utterance from LibriTTS-R, we use its text as input to CosyVoice2. The paired L2 sample serves as the LM prompt to impart the non-native accent, while the original L1 utterance itself serves as the Flow prompt to preserve the speaker's timbre.

The result of this process is a parallel L2-accented corpus aligned with the original LibriTTS-R data in content and speaker identity, ready for training our model.
Due to accentedness, the total duration is approximately $1.3 \times$ that of the native counterparts.

\subsection{Model Architecture}
\begin{figure}[htbp]
\centering
\includegraphics[width=0.46\textwidth]{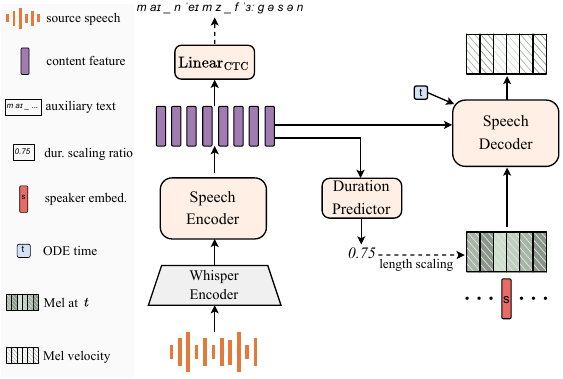}
\caption{CosyAccent architecture. It implicitly models rhythm for prosodic flexibility, while allowing the total duration to be either specified or predicted.}
\vspace{-0.3cm}
\label{fig:architecture}
\end{figure}

The architecture of CosyAccent is illustrated in Figure~\ref{fig:architecture}. It is a NAR system composed of four main modules: a speech encoder, a CTC projection head, a duration predictor, and a speech decoder.

\noindent\textbf{Speech encoder and content representations.}
The conversion process begins with the L2 source audio, which is first processed by a frozen Whisper$_\text{medium}$~\cite{radford2023robust} encoder frontend. The output is then passed to the Transformer speech encoder to extract high-level features. To ensure these features robustly represent linguistic content, we attach a linear projection head followed by a Connectionist Temporal Classification (CTC)~\cite{graves2006connectionist} loss to the encoder's output.


\begin{figure}[htbp]
\centering
\vspace{-0.4cm}
\includegraphics[width=0.39\textwidth]{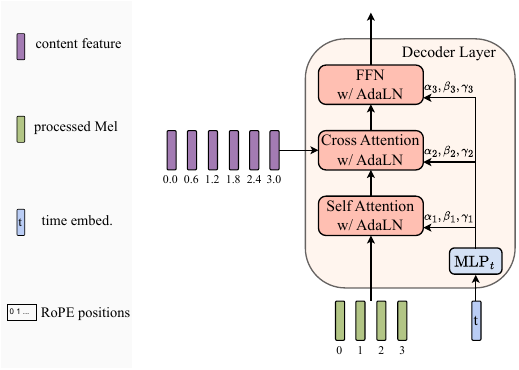}
\vspace{-0.3cm}
\caption{Speech decoder's alignment mechanism. Positional indices for the source content features are scaled to match the target's length, creating a coarse alignment within RoPE-based cross attention.}
\vspace{-0.3cm}
\label{fig:decoder}
\end{figure}

\noindent\textbf{Speech decoder}.
The DiT~\cite{peebles2023scalable} speech decoder is trained using flow matching~\cite{lipman2023flow} to generate the velocity of the Mel-spectrogram. As illustrated in Figure~\ref{fig:decoder}, each decoder layer comprises three modules: self-attention, cross-attention, and a feed-forward network (FFN). Each of these modules is followed by adaptive layer normalization (AdaLN), modulated by the time embedding. The encoder output, which provides the content representations, acts as the content condition and is incorporated through cross-attention.

\noindent\textbf{Total-duration control via position scaling}.
A key challenge in non-autoregressive conversion is aligning source and target sequences of different lengths. Our cross-attention decoder addresses this using Rotary Positional Encoding (RoPE)~\cite{su2024roformer} combined with a ``position scaling" technique, inspired by ARDiT TTS~\cite{liu2024autoregressive}.

Instead of using absolute indices for positional encoding, we normalize the positions of the source content features. As shown in Figure~\ref{fig:decoder}, the positional indices of the source encoder features are scaled so that their endpoint aligns perfectly with the endpoint of the target Mel-spectrogram's length. This technique establishes a coarse alignment between the source and target, ensuring the decoder correctly maps the start, middle, and end of the content regardless of the absolute length. Consequently, the model can robustly generate coherent output even when the target length is manually specified, a crucial feature for duration-constrained tasks like dubbing.

\noindent\textbf{Total-duration prediction}.
In accent normalization, the source speech has a known length that can be directly inherited for generation. This is a practical approach for applications like dubbing where duration must be strictly preserved. To provide flexibility for scenarios where the length can be modified without manual specification, we introduce a total-duration predictor. This module learns to predict a single \textit{total-duration scaling ratio}, defined as the target length divided by the source length, as depicted in Figure~\ref{fig:decoder}. For example, if an 8-frame input corresponds to a 6-frame target, the ratio is 0.75. During inference, this predicted ratio can be optionally used to determine the output length. The predictor is constructed with a DiT backbone followed by an attentive pooling layer and is trained via flow matching.

\noindent\textbf{Timbre Conditioning.}
While the frozen Whisper encoder~\cite{radford2023robust} provides robust linguistic features, it may suppress timbre information from the source audio. To ensure high-fidelity vocal mimicry, we augment the model with an explicit timbre condition by extracting a speaker embedding using a pre-trained speaker encoder. The total-duration predictor also uses the speaker embedding as input; this connection is omitted from the figure for clarity.

The decoder's velocity field is therefore conditioned on three inputs: the noisy sample $\mathbf{x}_t$ at time $t$, a sequence of content features $\mathbf{c}$, and a single speaker embedding vector $\mathbf{s}$. With parameterization $\theta$, the model's output is denoted as $v_\theta(\mathbf{x}_t, t, \mathbf{c}, \mathbf{s})$. To control the influence of these two distinct conditions, we employ a two-way Classifier-Free Guidance (CFG)~\cite{ho2021classifierfree} scheme in inference:
\begin{equation}
\begin{aligned}
    \bar{v}_\theta (\mathbf{x}_t, t, \mathbf{c}, \mathbf{s}) = & \ v_{\theta}(\mathbf{x}_t, t, \mathbf{c}, \mathbf{s}) \\
    & + w_1 (v_{\theta}(\mathbf{x}_t, t, \mathbf{c}, \mathbf{s}) - v_{\theta}(\mathbf{x}_t, t, \varnothing, \varnothing)) \\
    & + w_2 (v_{\theta}(\mathbf{x}_t, t, \mathbf{c}, \mathbf{s}) - v_{\theta}(\mathbf{x}_t, t, \varnothing, \mathbf{s}))
\end{aligned}
\label{eq:cfg}
\end{equation}
where $w_1$ and $w_2$ are guidance strengths, and $\varnothing$ denotes a dropped condition. The $w_1$ term provides general guidance, steering the generation away from the fully unconditional output. The $w_2$ term additionally strengthens the model's adherence to the linguistic content.

\section{Experimental Setup}
\label{sec:setup}

\subsection{Datasets}
Our experiments utilize several datasets. The native target corpus, LibriTTS-R ~\cite{koizumi2023librittsr}, is utilized by all systems. For training the baseline models, we use real-world L2 source speech from two datasets: the publicly available 20-hour L2-ARCTIC corpus~\cite{zhao2018l2arctic} and an internal 300-hour Chinese-accented English dataset. For our proposed method, we use the source-synthesized L2 accent dataset constructed in Section~\ref{subsec:data_construction}, which is derived from LibriTTS-R and L2-ARCTIC.

\subsection{Compared Systems}
We evaluate our model against two strong baselines:
\begin{itemize}
    \item \textbf{\textit{FramAN}}~\cite{bai2024diffusion}: A frame-to-frame flow-matching model.
    \item \textbf{\textit{TokAN}}~\cite{bai25_interspeech}: A model based on discrete SSL tokens. We test two modes: \textit{TokAN-1}, which predicts token durations directly, and \textit{TokAN-2}, which predicts with total-duration awareness and preserves the total duration.
\end{itemize}
Both baseline systems were trained on paired data consisting of real L2 speech (from L2-ARCTIC and the Chinese-accented set) and their corresponding L1 targets from LibriTTS-R.

For our proposed model, we also evaluate two configurations:
\begin{itemize}
    \item \textbf{\textit{CosyAccent-1}}: Uses the predicted total duration scaling ratio.
    \item \textbf{\textit{CosyAccent-2}}: Inherits the source total duration.
\end{itemize}
Crucially, CosyAccent is trained only on our source-synthesized dataset. This means that, unlike the baselines, our model is \textbf{never exposed to real L2-accented speech} during training.

For the CosyAccent architecture, we use the official Whisper-medium model~\cite{radford2023robust} as the frozen speech frontend and Resemblyzer\footnote{https://github.com/resemble-ai/Resemblyzer} for speaker embedding extraction. The final waveform is generated using the HiFTNet vocoder~\cite{li2023hiftnet} from CosyVoice2. During inference, the CFG weights $w_1$ and $w_2$ are both set to 1.0, and we use a 32-step Euler sampler for generation.

\begin{table*}[htb]
  \centering
  \caption{Evaluation results of the accent normalization systems}
  \vspace{-0.3cm}
  \small  
  \begin{tabular}{lcccccccc}
    \toprule
    \multirow{2}{*}{System} & \multirow{2}{*}{Source-length} & \multicolumn{3}{c}{Subjective}&\multicolumn{4}{c}{Objective} \\ \cmidrule(r){3-5}\cmidrule(r){6-9}
    & & NAT ($\uparrow$) & ACT ($\downarrow$) & SIM ($\uparrow$) & WER (\% $\downarrow$) & UTMOS ($\uparrow$) & SECS ($\uparrow$) & $\Delta\text{PPG}$ ($\downarrow$) \\
    \midrule
    Source & $\checkmark$ & 65.78{\tiny$\pm$2.18} & 50.45{\tiny$\pm$2.22} & - & 15.86 & 2.81 & - & 0.51 \\
    \midrule 
    FramAN~\cite{bai2024diffusion} & $\checkmark$ & 58.13{\tiny$\pm$2.19} & 44.08{\tiny$\pm$2.19} & $-$0.075 & 21.54 & 2.56 & 0.8065 & 0.49 \\
    TokAN-1~\cite{bai25_interspeech} & $\times$ & \underline{63.63}{\tiny$\pm$1.97} & {\bf 29.44}{\tiny$\pm$1.87} & $\phantom{-}${\bf 0.060} & 16.21 & 2.86 & \underline{0.8563} & {\bf 0.30} \\
    TokAN-2~\cite{bai25_interspeech} & $\checkmark$ & 57.25{\tiny$\pm$2.19} & 31.98{\tiny$\pm$2.00} & $-$0.027 & 16.71 & 2.76 & {\bf 0.8613} & \underline{0.30} \\
    CosyAccent-1 & $\times$ & {\bf 64.62}{\tiny$\pm$1.92} & \underline{31.04}{\tiny$\pm$1.91} & $\phantom{-}$\underline{0.033} & {\bf 12.96} & {\bf 3.04} & 0.8213 & 0.38 \\
    CosyAccent-2 & $\checkmark$ & 60.98{\tiny$\pm$2.05} & 35.19{\tiny$\pm$2.09} & $\phantom{-}$0.008 & \underline{13.26} & \underline{2.97} & 0.8291 & 0.37 \\
    \bottomrule
  \end{tabular}
\vspace{-0.5cm}
\label{tab:res}
\end{table*}

\subsection{Evaluation Data \& Metrics}

\noindent\textbf{Evaluation Set.}
Our test set is built from an extended L2-ARCTIC dataset~\cite{zhao2018l2arctic, kominek2004cmu}, covering seven accents: Arabic, Chinese, Hindi, Korean, Spanish, Vietnamese, and native American English. The set contains 80 sentences. This partitioning is consistent with our training data construction (Section~\ref{subsec:data_construction}).

\noindent\textbf{Subjective Evaluation.}
We conducted listening tests with 24 raters to assess three qualities: \textit{Naturalness (NAT)} and \textit{Accentedness (ACT)} were measured via MUSHRA tests. The native accent was excluded from the ACT evaluation. \textit{Speaker Similarity (SIM)} was measured via Best-Worst Scaling (BWS), with scores aggregated using a standard counting algorithm~\cite{ravillion2020comparison}: $(N_\textit{best} - N_\textit{worst}) / N_\textit{occurrence}$.

\noindent\textbf{Objective Evaluation.}
We use four objective metrics to assess conversion quality automatically. \textit{Intelligibility:} Word Error Rate (WER) from a native-only ASR model\footnote{https://huggingface.co/facebook/s2t-medium-librispeech-asr} to simulate listener perception. \textit{Naturalness:} The UTMOSv2 score\footnote{https://github.com/sarulab-speech/UTMOSv2} from a neural naturalness predictor. \textit{Timbre Preservation:} Speaker Encoding Cosine Similarity (SECS) using the accent-robust Resemblyzer. \textit{Accentedness Reduction:} The phonetic posteriorgram distance ($\Delta$PPG)~\cite{churchwell2024high}.

\vspace{-0.1cm}

\section{Results}
\label{sec:results}

\subsection{Comparison with Frame-to-Frame Baseline}
The subjective and objective evaluation results are presented in Table~\ref{tab:res}.
CosyAccent significantly outperforms the frame-to-frame baseline, FramAN, across all subjective and objective metrics. We attribute this broad superiority to our model's holistic spectrogram generation. In contrast, FramAN's reliance on explicit, frame-level pitch and energy predictors proves brittle, especially when encountering the unseen rhythmic patterns of L2 speech, leading to degraded naturalness and speaker similarity.

\vspace{-0.3cm}

\subsection{Comparison with Token-Based Baseline}

The comparison with the token-based \textit{TokAN} is more nuanced and reveals key trade-offs in accent conversion.

\noindent\textbf{Intelligibility/Content Preservation (WER).}
CosyAccent achieves significantly better content preservation, with a WER of 12.96\% compared to TokAN's 16.21\%. Our analysis of the generated samples suggests that TokAN performs an overly aggressive pronunciation modification. While this can aid in accent reduction, it frequently alters the underlying content, resulting in higher WER. As shown in Table~\ref{tab:accent_wise_wer}, CosyAccent's advantage holds consistently across nearly all accents, though performance is similar for Chinese and Vietnamese. This similarity may stem from shared prosodic traits (e.g., syllable-timed rhythm) and the fact that TokAN's training data included additional Chinese-accented speech.

\noindent\textbf{Naturalness and Speaker Similarity (NAT, UTMOS, SIM, SECS).}
In terms of speech quality, CosyAccent demonstrates superior naturalness over TokAN in both duration-preserving (\textit{-2}) and duration-predicting (\textit{-1}) modes, as confirmed by both subjective NAT and objective UTMOS scores. However, TokAN achieves higher speaker similarity (SECS). We hypothesize this is due to the fusion method: TokAN employs AdaLN to inject speaker information, a technique often more effective than the simple input concatenation. 
A notable discrepancy arises between the objective and subjective similarity scores: despite TokAN-2's superior objective SECS score, it is outperformed by CosyAccent on SIM. We attribute this to prosodic artifacts; the exaggerated rhythm in TokAN-2's output likely penalizes the human perception of speaker identity, even if the underlying vocal timbre is matched.

\noindent\textbf{Accentedness Reduction (ACT, $\Delta$PPG).}
In terms of accentedness reduction, CosyAccent performs on-par with TokAN, demonstrating the effectiveness of our source-synthesis approach. The slight performance difference can be attributed to the training data: while TokAN was trained on real L2-ARCTIC samples, CosyAccent was trained exclusively on our source-synthesized data. This domain gap highlights the challenge of bridging synthetic and real L2 speech, yet our results demonstrate that source-synthesized training data is not only viable but highly effective. The strong accentedness reduction achieved by CosyAccent validates our paradigm shift and eliminates the dependency on real L2 data collection.

\begin{table}[!htb]
  \centering
  \vspace{-0.2cm}
  \caption{Accent-wise WERs with native-only ASR}
  \vspace{-0.2cm}
  \label{tab:accent_wise_wer}
  \setlength{\tabcolsep}{5pt}  
  \small
  \resizebox{0.46\textwidth}{!}{
  \begin{tabular}{lccccccc}
    \toprule
    System & Zh & Hi & Vi & Ar & Es & Ko \\
    \midrule
    Source & 20.82 & 11.75 & 31.40 & 15.09 & 15.79 & 13.78 \\
    \midrule
    FramAN & 26.76 & 19.98 & 34.36 & 22.96 & 22.28 & 18.91 \\
    TokAN-1 & \underline{17.18} & 18.53 & \underline{24.51} & 17.25 & 16.68 & 13.81 \\
    TokAN-2 & 17.80 & 18.95 & 25.20 & 17.36 & 17.52 & 14.43 \\
    CosyAccent-1 & {\bf 16.95} & \underline{7.55} & 25.53 & {\bf 13.97} & {\bf 13.22} & {\bf 10.15} \\
    CosyAccent-2 & 17.77 & {\bf 7.52} & {\bf 24.41} & \underline{13.97} & \underline{14.40} & \underline{11.03} \\
    \bottomrule
  \end{tabular}
  }
\vspace{-0.6cm}
\end{table}

\subsection{Ablation Study}
We conduct an ablation study on CosyAccent-2, with results shown in Table~\ref{tab:ablation_res}. \textit{CTC auxiliary loss}: removing the CTC loss results in a significant increase in WER; this confirms that the auxiliary loss is crucial for guiding the model's content encoder and ensuring accurate content preservation during conversion. \textit{Speaker embeddings}: ablating the speaker embedding leads to a sharp drop in speaker similarity; this finding validates our hypothesis that the frozen Whisper encoder largely suppresses timbre information, making an explicit speaker condition essential in the current model design.
\textit{Content position scaling}: removing position scaling from the content features caused immediate training instability, leading to model collapse; this demonstrates that the technique is beneficial for stabilizing the training process of our proposed architecture.

\begin{table}[!htb]
  \centering
  \vspace{-0.3cm}
  \caption{Ablation results of CosyAccent-2}
  \vspace{-0.2cm}
  \label{tab:ablation_res}
  \setlength{\tabcolsep}{4pt}  
  \small
  \resizebox{0.45\textwidth}{!}{
  \begin{tabular}{lccc}
    \toprule
    System & WER ($\% \downarrow$) & SECS ($\uparrow$) & $\Delta\text{PPG}$ ($\downarrow$) \\
    \midrule
    Source & 15.86 & - & 0.51 \\
    \midrule
    CosyAccent-2 & {\bf 13.26} & 0.8291 & {\bf 0.37} \\
    \quad w/o CTC loss & 15.61 & {\bf 0.8324} & 0.41 \\
    \quad w/o speaker embedding & 13.51 & 0.6524 & 0.37 \\
    \quad w/o position scaling & \multicolumn{3}{c}{Model Collapse} \\
    \bottomrule
\end{tabular}
}
\vspace{-0.5cm}
\end{table}

\section{Conclusion, Limitations, \& Future Work}

In this paper, we introduced CosyAccent, a duration-controllable accent normalization model, and a novel source-synthesis data strategy. Our method largely reduces the need for real L2 training data by synthesizing the source speech and using authentic native speech as targets. This approach allows the model to avoid learning from TTS artifacts, leading to superior content preservation and naturalness. Key limitations include robustness to acoustic noise and control over paralinguistics, as the synthetic data is very clean. Future work will focus on these issues.

\vfill
\pagebreak

\section{Acknowledgement}
This work was supported by National Natural Science Foundation of China (Grant No. 62271432), Shenzhen Science and Technology Research Fund (Fundamental Research Key Project, Grant No. JCYJ20220818103001002), Program for Guangdong Introducing Innovative and Entrepreneurial Teams (Grant No. 2023ZT10X044), and Yangtze River Delta Science and Technology Innovation Community Joint Research Project (2024CSJGG1100).

\bibliographystyle{IEEEbib}
\bibliography{strings,refs}

\end{document}